\documentstyle[12pt]{article}
\catcode`@=11

\def\beqra{\begin{eqnarray}} \def\eeqra{\end{eqnarray}}
\def\beqast{\begin{eqnarray*}} \def\eeqast{\end{eqnarray*}}
\def\beq{\begin{equation}}      \def\eeq{\end{equation}}
\def\be{\begin{enumerate}}   \def\ee{\end{enumerate}}

\def\fo{\hbox{{1}\kern-.25em\hbox{l}}}

\def\utgp{Theory Group\\ Department of Physics \\ University of Texas
\\ Austin, Texas 78712}

\def\gam{\gamma}
\def\Gam{\Gamma}
\def\la{\lambda}
\def\eps{\epsilon}

\def\si{\sigma}

\def\al{\alpha}
\def\Tha{\Theta}
\def\tha{\theta}
\def\vphi{\varphi}
\def\del{\delta}
\def\Del{\Delta}
\def\ab{\alpha\beta}

\def\Om{\Omega}

\def\til{\tilde}

\def\eqv{\equiv}

\def\pa{\partial}

\def\lag{\langle}
\def\rag{\rangle}

\def\ul{\underline}

\def\nd{\noindent}
\def\ch{\@startsection{section}{1}{\z@}{-3ex plus-1ex minus-.2ex}%
        {2ex plus.2ex}{\large\sc}}

\def\cah{{\cal H}}

\def\cl{{\cal L}}
\def\cm{{\cal M}}

\def\cO{{\cal O}}
\def\cp{{\cal P}}
\def\cq{{\cal Q}}

\def\cv{{\cal{V}}}

\def\raisenot{\raise .5mm\hbox{/}}
\def\nota{\ \hbox{{$a$}\kern-.49em\hbox{/}}}
\def\notA{\hbox{{$A$}\kern-.54em\hbox{\raisenot}}}
\def\notb{\ \hbox{{$b$}\kern-.47em\hbox{/}}}
\def\notB{\ \hbox{{$B$}\kern-.60em\hbox{\raisenot}}}
\def\notc{\ \hbox{{$c$}\kern-.45em\hbox{/}}}
\def\notd{\ \hbox{{$d$}\kern-.53em\hbox{/}}}
\def\notbd{\ \hbox{{$D$}\kern-.61em\hbox{\raisenot}}} %big D
\def\note{\ \hbox{{$e$}\kern-.47em\hbox{/}}}
\def\notk{\ \hbox{{$k$}\kern-.51em\hbox{/}}}
\def\notp{\ \hbox{{$p$}\kern-.43em\hbox{/}}}
\def\notq{\ \hbox{{$q$}\kern-.47em\hbox{/}}}
\def\notW{\ \hbox{{$W$}\kern-.75em\hbox{\raisenot}}}
\def\notz{\ \hbox{{$Z$}\kern-.61em\hbox{\raisenot}}}

\def\notpa{\hbox{{$\partial$}\kern-.54em\hbox{\raisenot}}}

\def\llra{\relbar\joinrel\longrightarrow}

\def\7#1#2{\mathop{\null#2}\limits^{#1}}        % puts #1 atop #2
\def\5#1#2{\mathop{\null#2}\limits_{#1}}        % puts #1 beneath #2

\def\inbar{\vrule height1.5ex width.4pt depth0pt}
\def\IB{\relax{\rm I\kern-.18em B}}
\def\IC{\relax\leavevmode\hbox{\,$\inbar\kern-.3em{\rm C}$}}
\def\ID{\relax{\rm I\kern-.18em D}}
\def\IE{\relax{\rm I\kern-.18em E}}
\def\IF{\relax{\rm I\kern-.18em F}}
\def\IG{\relax\leavevmode\hbox{\,$\inbar\kern-.3em{\rm G}$}}
\def\IH{\relax{\rm I\kern-.18em H}}
\def\II{\relax{\rm I\kern-.18em I}}
\def\IK{\relax{\rm I\kern-.18em K}}
\def\IL{\relax{\rm I\kern-.18em L}}
\def\IM{\relax{\rm I\kern-.18em M}}
\def\IN{\relax{\rm I\kern-.18em N}}
\def\IO{\relax\leavevmode\hbox{\,$\inbar\kern-.3em{\rm O}$}}
\def\IP{\relax{\rm I\kern-.18em P}}
\def\IQ{\relax\leavevmode\hbox{\,$\inbar\kern-.3em{\rm Q}$}}
\def\IR{\relax{\rm I\kern-.18em R}}
\def\sed{\hbox{{\sf S}\kern-.4em\hbox{\sf S}}}
\def\ZZ{\relax{\sf Z\kern-.4em Z}}
\def\smIR{\hbox{{\footnotesize\rm I}\kern-.2em\hbox{\footnotesize\rm R}}}
\def\smIO{\ \hbox{{\footnotesize\rm I}\kern-.4em\hbox{\footnotesize\bf O}}}
\def\smIQ{\ \hbox{{\footnotesize\rm I}\kern-.5em\hbox{\footnotesize\bf Q}}}
\def\IGa{\relax{\rm I}\kern-.18em\Gamma}
\def\IPi{\relax{\rm I}\kern-.18em\Pi}
\def\IQt{\relax\leavevmode\hbox{$\kern.3em\inbar\kern-.3em\Theta$}}
\def\IOm{\relax\hbox{$\kern3.48pt\inbar\kern1.8pt\inbar\kern-5.28pt\Omega$}}

\def\ca#1{\relax\ifmmode {\cal#1} \else$\cal#1$\fi}     % Calligraphic next
\def\Sf#1{\relax\ifmmode\hbox{\sf#1}\else{\sf#1}\fi}    % San-Serif next
\def\fibby{\ifcase\@ptsize                      % redefines the Roman font
                \font\tenrm=cmfib8\or           % into the Fibonacchi font,
                \font\elvrm=cmfib8 scaled\magstephalf\or        % at the
                \font\twlrm=cmfib8 scaled\magstep1 \fi}         % @ptsize
\def\TeXey{\ifcase\@ptsize\or\or                % Instead of LaTeX's cmx12
                \font\twlrm=cmr10 scaled\magstep1       % takes 10pt fonts
                \font\twlmi=cmmi10 scaled\magstep1      % and magnifies
                \font\twlit=cmti10 scaled\magstep1      % by 120%, just
                \font\twlbf=cmbx10 scaled\magstep1\fi}  % as TeX does.

\def\ch{\@startsection{section}{1}{\z@}{-3ex plus-1ex minus-.2ex}%
        {2ex plus.2ex}{\large\sc}}
\def\sch{\@startsection{subsection}{2}{\z@}{-1.5ex plus-1ex minus-.2ex}%
        {1pt plus.2ex}{\sc}}
\def\ssch{\@startsection{subsubsection}{3}{\z@}{-1ex plus-1ex minus-.2ex}%
        {1pt plus.2ex}{\small\sc}}
\def\seceq{\@addtoreset{equation}{section}%     % Numbers Eq.s within Sect.s
        \def\theequation{\thesection.\arabic{equation}}}        % (Sect.Eq)

\def\con{\ifmmode \hbox{\bf*} \else{\bf*}\fi}   % conjugation
\def\scon{\ifmmode \hbox{\footnotesize\rm\bf*} \else{\footnotesize\rm\bf*}\fi}

\def\0#1{\relax\ifmmode\mathaccent"7017{#1}%    % puts a little circle atop,
        \else\accent23#1\relax\fi}              % as a halo of a saint

\def\haf{\frac{1}{2}}

\def\place#1#2#3{\vbox to0pt{\kern-\parskip\kern-7pt
                             \kern-#2truein\hbox{\kern#1truein #3}
                             \vss}\nointerlineskip}

\def\illustration #1 by #2 (#3){\vbox to #2{\hrule width #1 height 0pt depth
0pt
                                       \vfill\special{illustration #3}}}

\def\scaledillustration #1 by #2 (#3 scaled #4){{\dimen0=#1 \dimen1=#2
           \divide\dimen0 by 1000 \multiply\dimen0 by #4
            \divide\dimen1 by 1000 \multiply\dimen1 by #4
            \illustration \dimen0 by \dimen1 (#3 scaled #4)}}

\catcode`@=12

\begin{document}
\renewcommand\theequation{\thesection.\arabic{equation}}
\def\cq{{\cal Q}}

%date:     Mon, 27 Jan 92 15:15 CST

\vspace*{24pt}
\begin{center}
{\large{\bf First Reduce or First Quantize? \\
A Lagrangian Approach and \\[3pt]
Application to Coset Spaces}}

\vspace{36pt}
 C.R. Ord\'o\~nez

\vspace{6pt}
{\it \utgp}

\vspace{24pt}
 J.M. Pons

\vspace{6pt}
{\it Center for Relativity \\
Department of Physics \\
University of Texas \\
Austin, Texas 78712}\\

\vspace{24pt}

\ul{ABSTRACT}

\end{center}

\vspace{18pt}
A Lagrangian treatment of the quantization of first class Hamiltonian systems
with constraints and Hamiltonian linear and quadratic in the momenta
respectively is performed.  The ``first reduce and then quantize'' and the
``first quantize and then reduce'' (Dirac's) methods are compared.  A new
source of ambiguities in this latter approach is revealed and its relevance on
issues concerning self-consistency and equivalence with the ``first reduce''
method is emphasized.  One of our main results is the relation between the
propagator obtained {\it \`a la Dirac} and the propagator in the full space,
eq.
(5.25).As an application of the formalism developed,
quantization on coset spaces of compact Lie groups is presented.  In this case
it is shown that a natural selection of a Dirac quantization allows for full
self-consistency and equivalence.  Finally, the specific case of the propagator
on a two-dimensional sphere $S^2$ viewed as the coset space $SU(2)/U(1)$ is
worked out.

\vfill

\pagebreak

\setcounter{page}{1}

\baselineskip=24pt
\section{Introduction}

As is well known, the quantization of constraint systems is plagued with
many ambiguities and difficulties which are added to those encountered when
dealing with regular systems
(ordering problems, Groenwald-van Hove obstruction  \cite{re:GS 84}, etc.).
In particular, in the framework of Dirac quantization \cite{re:D 50}, the
preservation of the first class nature of the Hamiltonian and the constraints
at
the operator level is a highly non-trivial issue.
Another important aspect in the quantization of constrained systems is
that of the equivalence between the two standard procedures:~~
a) Dirac's method of quantizing the ``entire'' system
(i.e.,  \ including gauge variables) and obtaining physical states as
those annihilated by the operator version of the constraints, and
{}~~b)  first reducing the classical variables by solving the constraints,
and then quantizing as a regular system.  These procedures are usually referred
to as ``quantize first and then reduce'' and
``reduce first and then quantize'' respectively.  There seems to be no
a-priori general principle that guarantees that they agree with each other, let
alone that selects one procedure over the other from the physical  point of
view \cite{re:f 1}.  In fact, some of these issues have recently been
discussed in the  context of Chern-Simons topological field theories
\cite{re:CI 90},\cite{re:GM 89},\cite{re:JA 89},\cite{re:GZ 90}, which
exemplifies the level of subtlety involved in this kind of problems.

Kucha\v r, in a beautiful series of papers  \cite{re:KU 86} , has discussed
these  problems in the case of quadratic Hamiltonians and linear constraints.
(Recently Kucha\v r and H\'aj\'\i\v cek have also studied the case of
parametrized theories with quadratic constraints  \cite{re:HK 90}.)
He finds that, in order to achieve equivalence, one is forced to give
up, in the general case, the full Hilbert space structure of the entire
system.
Hence, in a strict sense, in Dirac's method one is not
``quantizing first'' and then finding physical states, since the entire
space does not admit a quantum interpretation.
Following Kucha\v r's work, McMullan et~al. \cite{re:Mc 89} have used the BRST
approach of Batalin, Fradkin, and Vilkovisky \cite{re:BF 77} to address the
same
problems. They were able to implement the ``quantize first'' procedure, while
keeping the equivalence with the ``reduce first'' method, by means of
the introduction of ghost variables. Concurrently, several other groups
\cite{re:GK 91} have also investigated a wide variety of aspects which arise in
the study of  the equivalence of these two methods.

The treatment of these problems has been done mostly with canonical (operator)
quantization. However, similar issues like the relation between reduced
and covariant path integrals (e.g. Faddeev-Popov) also appear. We have
investigated   these issues in a series of publications \cite{re:OP}
\cite{re:OPT} \cite{re:OPPT} and
have proven equivalence in this case (see also \cite{re:FS}).
[6~
Let us now return to the operator formulation.

The work of the groups mentioned above has been mostly done within the
Hamiltonian framework.
By doing this, they may have overlooked a potential new source of
ambiguities stemming from the fact that the Hamiltonian for a classical
constrained system is not uniquely defined off the constraint surface, which is
going to have an impact upon its quantization.
For this reason, in order to tackle this type of problems, we have
chosen to start with a Lagrangian framework (section 2).
Beyond its well known advantages---explicit display of symmetries being
perhaps the most commonly quoted---this formalism enables us to encode
in a single function all the features of the constrained system,
including, of course, the constraints.  We then proceed with the Hamiltonian
formulation (section 3), and immediately afterwards we present the two types of
quantizations mentioned above (sections 4 and 5).  This is followed (section
6) by a discussion of the subtleties and ambiguities present in the ``quantize
first''  approach which, to our knowledge, have not been addressed before.
Here we also make some remarks on the relevance of the gauge group in this
kind of questions.  On section 7 we apply our general setting to the case of
quantization on coset spaces of compact Lie groups and then go on to the
special case of $S^2\approx SU(2)/U(1)$ (section 8).
Finally, we present conclusions and outlook in section 9.

\section{The Lagrangian setting}
\setcounter{equation}{0}

We are interested in a model with only first class constraints
(in phase space, $T^*{\cal Q}$, where ${\cal Q}$ is the configuration space
manifold), all of which should appear as primary constraints in order for them
to generate independent gauge transformations.
This means that also the Hamiltonian has to be first class, i.e.,
its Poisson brackets with all the constraints have to vanish on the
constraint surface.
 There is an easy way to cast this information in the Lagrangian
formalism.  In this paper we will consider Lagrangians of the form:
\begin{equation}
L={1\over 2}G_{AB}\dot Q^A\dot Q^B-V, \label{2.1}
\end{equation}
where $G_{AB}$ and $V$ are functions of configuration space variables
$Q^A$,  $A=1,\ldots,N$.
 $G_{AB}$~is a singular metric tensor of rank $n< N$ \cite{re:f 2}.

 The standard construction of the
first generation of velocity space $(T{\cal Q})$ constraints is as follows. The
equation of motion derived from (\ref{2.1}) is \cite{re:f 3}
\begin{equation} G_{AB}\ddot
Q^B=\alpha_A, \label{2.2}
\end{equation}
where
\begin{eqnarray}
\alpha_A&=&{\partial L\over\partial Q^A}
          -{\partial L\over\partial Q^B\partial\dot Q^A}
                     \dot Q^B\nonumber\\[\medskipamount]
&=&{1\over 2}G_{BC,A}\dot Q^B\dot Q^C
            -G_{AC,B}\dot Q^B\dot Q^C-V_{,A}. \label{2.3}
\end{eqnarray}

If $U^A(Q)$ is a null vector of $G_{AB}$, i.e., $G_{AB}U^B=0$
identically, then the constraint surface corresponding to the first step
of the stabilization algorithm is obtained by requiring \cite{re:f 4}
\begin{equation}
U^A\alpha_A\approx 0 \label{2.4}
\end{equation}
 for every null vector of the metric.

In order for (\ref{2.4}) not to select a constraint surface, this
relation has to be satisfied {\it identically}, i.e.,
its solution has to be the entire velocity space \cite{re:f 5}.
This leads to
\begin{eqnarray} U^C\Gamma_{CAB}&=&0\,,
\label{2.5}\\[\medskipamount]
\noalign{\noindent\rm and\medskip} U^CV_{,C}&=&0, \label{2.6}
\end{eqnarray}
where
\begin{equation}
\Gamma_{CAB}=-G_{AB,C}+G_{CB,A}+G_{CA,B}. \label{2.7}
\end{equation}
Using $G_{AB}U^B=0$, (\ref{2.5}) becomes
\begin{equation}
U^CG_{AB,C}+G_{AC}U^C_{,B}+G_{CB}U^C_{,A}\equiv
\left({\cal L}_{\hat U}G\right)_{AB}=0 \label{2.8}
\end{equation}
with
\begin{equation}
\hat U=U^A(Q){\partial\over\partial Q^A}. \label{2.9}
\end{equation}
In (\ref{2.8}) we have used the standard notation for the Lie
derivative.

Clearly, in this case there cannot be any further constraints.
Thus, the conditions which the Lagrangian (\ref{2.1}) has to satisfy in
order for it to guarantee a first class Hamiltonian structure are:

\begin{enumerate}
\item[a)] every null vector of the degenerate metric tensor $G$, considered
as a vector field in configuration space, also has to be a Killing
vector for it, and
\item[b)] every such null vector has to be tangent to the equipotential
surfaces of the potential~$V$.
\end{enumerate}

At this point, the following comment about the null vectors of $G$ is in
order.
The set of null vectors is closed under the Lie bracket.
Indeed, let $\hat U$ and $\hat V$ be two such vectors, i.e., \cite{re:f 6}
\begin{equation} i_{\hat U}G=i_{\hat V}G=0. \label{2.10}
\end{equation}
As we saw, they also have to be Killing vectors. Hence,using (\ref{2.8}) the
following equalities hold:
\begin{equation}
0={\cal L}_{\hat U}\left(i_{\hat V}G\right)
 =i_{\left({\cal L}_{\hat U}\hat V\right)}G
 =i_{[\hat U ,\hat V]}G. \label{2.11}
\end{equation}
Therefore, $[\hat U,\hat V]$ is a null vector.

If we consider a basis for the space of null vectors,
$\left\{\hat U_\alpha,\alpha=1,\ldots,k=N-n\right\}$
(this means that any null vector can be written as a linear combination
of the vectors in this basis, with its coefficients being functions of
configuration space variables), this closure property is written
\begin{equation}
\left[\hat U_\alpha,\hat U_\beta\right]=
C^\gamma_{\alpha\beta}(Q)\hat U_\gamma. \label{2.12}
\end{equation}
As we will see later this relation prefigures the first class character
of the constraints that appear in the Hamiltonian formulation.

\section{Hamiltonian setting}
\setcounter{equation}{0}

Now we will explicitly see how our previous scheme is realized in the
Hamiltonian framework.
The momenta are defined as usual by the legendre map $P=\pa L\big/\pa\dot Q$.
\begin{equation}
P_A=G_{AB}\dot Q^B. \label{3.1}
\end{equation}
This immediately implies the following set of primary constraints
\begin{equation}
\varphi_\alpha\equiv U^A_\alpha P_A \approx 0, \al = 1,\ldots, k\;. \label{3.2}
\end{equation}
Since the rank of the metric $G$ is $n$, (\ref{3.2}) is the full set of
primary constraints.
Equation (2.12)~shows that they are first class under Poisson bracket:
\begin{equation}
\left\{\varphi_\alpha,\varphi_\beta\right\}=
-C^\gamma_{\alpha\beta}\varphi_\gamma. \label{3.3}
\end{equation}
As mentioned earlier, these constraints generate gauge transformations.
For any trajectory $\eta=(Q^A(t), \;P_A(t))$ in $ T^*{\cal Q}$,
\beq
\del\, \eta = \eps^\al(t)\left\{\eta,\; \varphi_\alpha\right\},
  \label{3.4}
\eeq
where $\eps^\al(t)$ are $k$ arbitrary infinitesimal functions of $t$ which
describe $k$ independent gauge transformations.  For fixed $t$ these turn
into infinitesimal symmetry transformations in $T^*\cq$.
In particular, for the coordinates of
configuration space
\beq
\delta  Q^A= \eps^\al\left\{Q^A,\varphi_\alpha\right\}
  = \eps^\al U_\al^A\,, \label{3.5}
\eeq
where $\eps^\al$ are infinitesimal parameters.
Hence the vector fields $\hat U_\alpha$ generate symmetry transformations
in configuration space.  This means that configuration space $\cm$ is divided
into orbits under the action of these vector fields, and that the physical
configuration space is the quotient of the original one by the orbits.

Let us now set up the dynamics on $T^*\cq$.
The energy function in velocity space is
\renewcommand\theequation{\thesection.\arabic{equation}}
\begin{eqnarray}
E&=&{\partial L\over\partial\dot Q^A}
\dot Q^A-L\nonumber\\[\medskipamount]
&=&{1\over 2}G_{AB}\dot Q^A\dot Q^B+V. \label{3.6}
\end{eqnarray}
This function has to be the pullback under the Legendre mapping  of some
function in phase space, which, up to functions that vanish on the constraint
surface -- i.e., having zero pullback -- has the following general form:
\begin{equation}
H={1\over 2}M^{AB}(Q)P_AP_B+V. \label{3.7}
\end{equation}
This is, of course, the Hamiltonian function.
Notice that it is only uniquely defined on the constraint surface.
This fact is reflected in an ambiguity in~$M^{AB}$.
Indeed, by construction, the only requirement that it has to satisfy is
that ($M$~being symmetric)
\begin{eqnarray}
M^{AB}G_{AC}G_{BD}\dot Q^C\dot Q^D&=&
G_{CD}\dot Q^C\dot Q^D, \label{3.8}\\[\medskipamount]
\noalign{\noindent\rm i.e.,\medskip}
M^{AB}G_{AC}G_{BD}&=&G_{CD}. \label{3.9}
\end{eqnarray}
Equation~(\ref{3.9}) displays the ambiguity very clearly.
Any transformation of the form
\begin{equation}
M^{AB}\to M'{}^{AB}=M^{AB}+\lambda^{A\alpha}U^B_\alpha
                          +\lambda^{B\alpha}U^A_\alpha \label{3.10}
\end{equation}
with arbitrary $\lambda^{A\alpha}$ will define another Hamiltonian which
will have the same energy function in velocity space.
 This lack of uniqueness in $M^{AB}$ allows for the possibility of
working with a non-singular matrix which may well be used as a metric
tensor in configuration space.
Nevertheless, notice that this assignment is {\it highly ambiguous}.
This is to say, a first class Hamiltonian system of the type we are
considering {\it does not\/} endow configuration space with a
{\it unique\/} non-singular metric structure.
This is the point mentioned earlier which has not been taken into
account in previous work.
We would like to emphasize the importance of these remarks in the light
of the problems one faces when quantizing these systems.
Recall that we are to try to preserve the first class nature of the
constraints, and to check equivalence between the
``quantize and then reduce'' and the
``reduce and then quantize'' approaches.
This freedom in the choice of a metric structure in configuration space
will play an important role in our quantization program.

Finally, we would like to point out the generality of our construction.  It can
be shown that given a Hamiltonian of the form (\ref{3.7}), and a set of the
first class constraints (\ref{3.2}), there always exists a Lagrangian of the
form (\ref{2.1}) which gives rise to the Hamiltonian and the constraints.  For
details see appendix.

\section{Reduce first and then quantize}
\setcounter{equation}{0}

We now come to the description of the physical (reduced) configuration
space and the quantum theory on it.  At this level we may proceed from either
the Hamiltonian or the Lagrangian formulation.  As we will show, one obtains
the same results.  Let us then begin with the Hamiltonian version. There is a
natural way of endowing $\cal M$ with a non-singular metric structure. Consider
the following contravariant tensor field associated to the kinetic term in the
Hamiltonian
\begin{equation}
M^{AB}{\partial\over\partial Q^A}\otimes
      {\partial\over\partial Q^B}. \label{4.1}
\end{equation}
The projection
\begin{equation}
\pi:{\cal Q}\to\cal M \label{4.2}
\end{equation}
allows us to assign to (\ref{4.1}) a contravariant vector field in
$\cal M$, as long as the following ``projectability'' condition is
fulfilled:
\begin{eqnarray}
\hat U_\alpha\left(M^{AB}{\partial f\over\partial Q^A}
                   \,{\partial g\over\partial Q^B}\right) &=&0,
\qquad\alpha=1,\ldots,k \label{4.3}\\[\medskipamount]
\noalign{\noindent\rm for any functions $f$, $g$ of configuration
variables $Q^A$ such that\medskip}
\hat U_\alpha f=
\hat U_\alpha g&=&0,\qquad\alpha=1,\ldots,k\,. \label{4.4}
\end{eqnarray}
This condition can be recast as
\begin{equation}
\left({\cal L}_{\hat U_\alpha}M\right)^{AB}G_{AC}G_{BD}=0 \label{4.5}
\end{equation}
which is a trivial consequence of (\ref{3.9}).
Hence we have to our avail this contravariant tensor field in~$\cal M$.

In order to gain further insight into the structure of this tensor
field, we now introduce coordinates $q^a$, $a=1,\ldots,n,$ on~$\cal M$.
Then the projection $\pi$ is described by the functions $q^a(Q)$ that
satisfy $\hat U_\alpha q^a=0$; $\alpha=1,\ldots ,k$, $a=1,\ldots,n$.

In these coordinates the components of the new tensor field in $\cal M$
are
\begin{equation}
\tilde g^{ab}=M^{AB}{\partial q^a\over\partial Q^A}
                  \,{\partial q^b\over\partial Q^B}\,. \label{4.6}
\end{equation}
The projectability condition shows that $\tilde g^{ab}$ only depends on
the physical coordinates~$q^a$.

Notice that $\tilde g^{ab}$ so defined is invariant under changes of $M$
of the type (\ref{3.10}).
Moreover, these are the {\it only\/} changes that leave $\tilde g^{ab}$
invariant.
 This result has the direct physical consequence that
{\it the ambiguities present in the Hamiltonian framework play no role
in the reduce first approach to quantization}.  Hence the physical phase space
$T^*\cm$ is so naturally endowed with a Hamiltonian

\beq
h=\haf\;\tilde g^{ab} p_a \,p_b +V~. \label{4.7}
\eeq
Now we will show that $\tilde g$ is non-singular.  In order to prove this
it is convenient to work with a
coordinate system in $\cq$ adapted to the orbit structure.
This is achieved by adding to the physical coordinates $q^a$,
$a=1,\ldots ,n,$ which label the orbits, a new set of functions
$q^\alpha$, $\alpha=1,\ldots ,k$ such that
$\det\left|\hat U_\alpha\hat q^\beta\right|\not=0$
(these new ``gauge'' coordinates parametrize the orbits.
The entire set will henceforth be termed ``adapted coordinate system'').

In this system the Lagrangian metric $G_{AB}$ and the null vectors are
written
\beq
G=\left(\begin{array}{cc}g_{ab}&g_{a\beta}\\[\medskipamount]
g_{\alpha b}&g_{\alpha\beta}\end{array}\right)
\label{4.8}
\eeq
and
\beq
\hat U_\alpha = U^\beta_\alpha\left(q^a,q^\gamma\right)
\frac{\partial}{\partial q^{\beta}}\,, \label{4.9}
\eeq
respectively.

The condition that these vectors are null vectors for this metric
immediately leads to
\begin{equation}
g_{\alpha b}=g_{a\beta}=g_{\alpha\beta}=0 \label{4.10}
\end{equation}
and the Killing condition for these vectors becomes:
\begin{equation}
\hat U_\alpha\left(g_{ab}\right)=0,\qquad
\alpha=1,\ldots,k\,. \label{4.11}
\end{equation}
This last result guarantees that $g_{ab}$ only depends on the physical
coordinates $q^a$, $a=1,\ldots ,n$.
Equations~(\ref{4.8}) and (\ref{4.10}) display the non-singularity of
$g_{ab}$ because the rank of $G$ is~$n$.

Equation~(\ref{3.9}) now becomes
\begin{equation}
M=\left(\begin{array}{cc}g^{ab}&m^{a\beta}\\[\medskipamount]
m^{\alpha b}&m^{\alpha\beta}\end{array}\right), \label{4.12}
\end{equation}
where $g^{ab}=\left(g_{ab}\right)^{-1}$, and the remaining components of
$M$ are arbitrary.

Finally, the projection $\pi$ to $\cal M$ of the contravariant tensor
(\ref{4.1}) leads to the identification $\tilde g^{ab}=g^{ab}$.
This proves that the contravariant metric tensor defined in $\cal M$ is
non-singular.  The Hamiltonian formulation on $T^*\cm$ is now completed.

Let us next proceed with the Lagrangian version.  Taking advantage of the form
of $G$ in the adapted coordinate system, it is immediate to see that there is
a unique function $\ell$ on $T\cm$ such that its pullback $\pi^{\prime*}\ell$
under the derivative mapping $\pi'$ induced by the projection $\pi$, i.e.,
\beq
\pi'\;:\; T\cq ~ \llra ~ T\cm \label{4.13}
\eeq
is just the original Lagrangian $L$:
\beq
\ell=\haf~ g_{ab}\,\dot q^a\, \dot q^b - V~~,~~\pi^{\prime*}\ell=L\;.
\label{4.14} \eeq
Summarizing: \ we have found the Hamiltonian and the Lagrangian for the reduced
system.  They are obviously related by a Legendre transformation.  In the
construction of these functions it now becomes quite clear that the metric
$g_{ab}$ is the one to be used to define the measure on $\cm$.  From this point
on, we can proceed within the standard quantization scheme for regular systems.
(See \cite{re:KU 86} and \cite{re:DW 57} for details).  Notice that in this
geometrical framework no explicit gauge fixing was performed.

The inner product that defines the Hilbert space $\cah_r=\cl^2(\cm,|g|^{1/2})$
is defined as
\begin{equation}
\left<\psi_1|\psi_2\right>=\int_{\cal M}d^nq\,|g|^{\haf}
  \psi^*_1(q)\psi_2(q)\,, \label{4.15}
\end{equation}
where $|g|$ is the determinant of $g_{ab}$.
The dynamical evolution (Schrodinger equation) will be given by the
Hamiltonian operator
\begin{eqnarray}
\hat h&=&-\,{1\over 2}~\Delta_g+V, \label{4.16}\\[\medskipamount]
\noalign{\noindent\rm where $\Delta_g$ is the Laplace-Beltrami
operator\medskip}
\Delta_g&=&|g|^{-\haf}\frac{\partial}{\partial q^a}
      |g|^{\haf}g^{ab}\frac{\partial}{\partial q^b}\,. \label{4.17}
\end{eqnarray}
In Kucha\v r's words, (\ref{4.16}) implements the
``principle of minimal coupling'';
curvature terms  are not included.

The configuration variables $q^a$ become multiplicative operators as in
the usual case, whereas, in order to retain hermiticity, the canonical
momenta $p_a$ are quantized as follows:
\begin{equation}
  p_a\to-i\left({\partial\over\partial q^a}+
{1\over 2}~|g|^{-\haf}|g|^{\haf}_{,a}\right). \label{4.18}
\end{equation}
In the next section we reverse the order of things and proceed to
``quantize first and then reduce''.

\section{Quantize first and then reduce, Dirac's Method}
\setcounter{equation}{0}

As we discussed in Section~3 we always have an enormous freedom to
choose a non-singular metric $M$ with which we can immediately write
down an inner product.
\begin{equation}
\left<\psi_1\,|\psi_2\right>=\int_Q
d^NQ\,|M|^{\haf}\psi^*_1(Q)\psi_2(Q). \label{5.1}
\end{equation}
This inner product defines the Hilbert space
${\cal H}={\cal L}^2\left({\cal Q},|M|^{\haf}\right)$.\footnote{Here
$|M|=\det\; M_{AB},\qquad M_{AB}=(M^{AB})^{-1}$.}
At this point we will assume that $M$ allows for a consistent
quantization of our classical first class Hamiltonian system, in the
sense defined in Section~1.
That is to say that the quantum operators associated with the classical
constraint and the Hamiltonian still form a first class system with
respect to the commutator algebra.
For a further discussion of this assumption we refer to the next
section.

The operator assignment for the classical constraints is dictated by the
requirement that the physical wave function only depend upon physical
variables, i.e., variables describing~$\cal M$:
\begin{equation}
 \vphi_\al \llra \hat\vphi_\al \eqv \hat U_\al=U^A_\al\; \frac{\pa}{\pa Q^A}
\label{5.2}
\end{equation}
Notice that equation (\ref{2.12}) guarantees that the classical constraints
$\vphi_\al $ are realized as a set of first class quantum operators.  The
Hamiltonian in $\cal H$ is chosen as in Section~4, i.e.,
\begin{eqnarray}
\hat H&=&-{1\over 2}~\Delta_M+V, \label{5.3}\\[\medskipamount]
\noalign{\noindent\rm where\medskip}
\Delta_M&=&|M|^{-\haf}{\partial\over\partial Q^A}~
      |M|^{\haf}M^{AB}{\partial\over\partial Q^B} \label{5.4}
\end{eqnarray}
and it is assumed to satisfy the first class condition:
\begin{equation}
\left[\hat U_\alpha,\hat H\right]=
\lambda^\beta_\alpha(Q)\hat U_\beta. \label{5.5}
\end{equation}
The physical Hilbert space ${\cal H}_p$ in this framework is then
obtained as the subspace of $\cal H$ defined by those states which
satisfy
\begin{equation}
\hat U_\alpha\left|\mbox{Phys}\right>=0,
 \qquad\alpha=1,\ldots ,k\,. \label{5.6}
\end{equation}
In wave function language
\begin{equation}
\hat U_\alpha\left<Q\mid\psi\right>\eqv \hat U_\alpha\psi(Q)=0. \label{5.7}
\end{equation}
This is equivalent to
\begin{equation}
\psi(Q)=\tilde\psi\left(f^a(Q)\right), \label{5.8}
\end{equation}
where $q^a=f^a(Q)$ defines explicitly the projection $\pi$, equation
(\ref{4.2}). Physical position kets $\left|q^a\right>_{\rm Ph}$ are those
states
in $\cal H$ defined by
\begin{equation}
_{\rm Ph\!}\left<q^a\mid\psi\right>
=\tilde\psi\left(q^a\right), \label{5.9}
\end{equation}
with $|\psi\rag$ satisfying (\ref{5.6}).
It is convenient to expand these states in terms of the position states
$\left|Q\right>$ in~$\cal H$.
For this purpose consider the following string of identities, valid for
arbitrary~$\tilde\psi$:
\begin{eqnarray}
\tilde\psi\left(q^a\right)={}_{\rm Ph\!}\left<q^a\mid\psi\right>
    &=&\int d^NQ\,|M|^{\haf}\,_{\rm Ph\!}\left<q^a\mid Q\right>
\left<Q\mid\psi\right>\nonumber\\[\medskipamount]
&=&\int d^NQ\,|M|^{\haf}\,
_{\rm Ph\!}\left<q^a\mid Q\right>\psi(Q)\nonumber\\[\medskipamount]
&=&\int d^NQ\,|M|^{\haf}\,_{\rm Ph\!}\left<q^a\mid Q\right>
\tilde\psi\left(f^a(Q)\right) \label{5.10}
\end{eqnarray}
which implies that
\begin{eqnarray}
_{\rm Ph\!}\left<q^a \mid Q\right>&=&{1\over\mu\left(q^a\right)}\delta^n
\left(q^a-f^a(Q)\right), \label{5.11}\\[\medskipamount]
\noalign{\noindent\rm where\medskip}
\mu(q^a)&=&\int d^NQ\,|M|^{\haf}\delta^n
\left(q^a-f^a(Q)\right). \label{5.12}
\end{eqnarray}

Notice that (\ref{5.11}) requires that $\mu(q^a)$ be finite at every
point~$q^a$.
 This is equivalent to requiring that the physical wave functions $\psi$
be normalizable in $\cal H$, since they are constant along each orbit.
Hence, in a rigorous sense, ``first  quantize'' approach
is only applicable in the cases where this condition is satisfied.
Of course, one can be more practical, and carry out the Dirac
construction in a more formal fashion ignoring details of normalization
in~${\cal H}$.
In our paper we are going to assume that $\mu\left(q^a\right)$ is finite
at every~$q^a$.

It will prove convenient to rewrite $\mu\left(q^a\right)$ without the
delta function.
For this purpose, let us use the adapted coordinate system
($q^a$,~$q^\alpha$) \cite{re:f 7}.  Consider the following identity \cite{re:f
8}. \begin{eqnarray}
1&=&\int d^kq~\delta^k\left(q^\alpha-f^\alpha(Q)\right),
\label{5.13}\\[\medskipamount]
\noalign{\noindent\rm where the functions
$f^\alpha$ give the explicit realization of the coordinate change.
Inserting (\ref{5.13}) into   (\ref{5.12}) we obtain\medskip}
\mu\left(q^a\right) &=& \int d^kq\int d^NQ\,|M|^{\haf}\delta^n
   \left(q^a-f^a(Q)\right)\delta^k
   \left(q^\alpha-f^\alpha(Q)\right)\nonumber\\[\medskipamount]
&=&\int d^kq\int d^NQ\,|M|^{\haf}\delta^N
\left(q-f(Q)\right)\nonumber\\[\medskipamount]
&=&\int d^kq\int d^NQ\,|M|^{\haf}
{1\over\,|\partial q/\partial Q|\,}\delta^N
\left(Q-Q(q)\right)\nonumber\\[\medskipamount]
&=&\int d^kq|M|^{\haf}{1\over\,|\partial q/\partial Q|\,}
  =\int d^kq|m|^{\haf}, \label{5.14}
\end{eqnarray}
where $|m|$ is the determinant of the metric in the adapted coordinate
system.

In \cite{re:OP} we proved the following important factorization property of
$m$:
in adapted coordinates,
\beq
|m|=|g_{ab}|\;|m_{\ab}|\eqv|g|\;|m_{\ab}|  \label{5.A}
\eeq
where $g_{ab}(q^a)$ is the ``physical'' metric of section 4.  This implies
that eq. (5.14) can also be written as
\begin{eqnarray}
\mu(q^a) &=& \int\,d^kq\;|g|^{1/2}\;|m_{\ab}|^{1/2}=|g|^{1/2}\; \int\;
d^kq\,|m_{\ab}|^{1/2} \nonumber \\
&=& |g|^{1/2}\; {\cv(q^a})\,.   \label{5.B}
\end{eqnarray}
$\cv(q^a)$ is naturally identified as the volume of the orbit labeled by $q^a$
(recall the $q^\al$'s are ``gauge'' variables).  Notice that since the
hipotesis $\mu(q^a)$ if finite, so is $\cv(q^a)$~~ (or, as this formula shows,
one can demand finiteness of $\cv(g^q)$ instead from the outset).

Using (\ref{5.11}) we can easily compute the inner product between any
two physical position kets:

\begin{eqnarray}
{}~_{\rm Ph\!}\lag q^a_1\mid q^a_2\rag_{\rm Ph}&=&\int d^NQ\,|M|^{\haf}\,
_{\rm Ph\!}\lag q^a_1\mid Q\rag
\left<Q\mid q^a_2\right>_{\rm Ph}\nonumber\\[\medskipamount]
&=&\int d^NQ\,|M|^{\haf}{1\over\mu\left(q^a_1\right)}\delta^n
\left(q^a_1-f^a(Q)\right){1\over\mu\left(q^a_2\right)}\delta^n
\left(q^a_2-f^a(Q)\right)\nonumber\\[\medskipamount]
&=&\delta^n\left(q^a_1-q^a_2\right){1\over\mu\left(q^a_1\right)}\,
 {1\over\mu\left(q^a_2\right)}\int d^NQ\,
|M|^{\haf} \,\delta^n\left(q^a_1-f^a(Q)\right) \nonumber\\[\medskipamount]
&=&\frac{1}{\mu\left(q^a_1\right)}\delta^n
\left(q^a_1-q^a_2\right). \label{5.15}
\end{eqnarray}

\nd
Equation~(\ref{5.15}) identifies $\mu\left(q^a\right)$ as the measure
on~$\cal M$.
This can also be seen directly if we compute the inner product of two
physical states $\left|\psi\right>$, $\left|\chi\right>$ in~${\cal H}_p$:
\begin{eqnarray}
\left<\psi\mid \chi\right>&=&\int d^NQ\,|M|^{\haf}\psi^*(Q)
\chi(Q)\nonumber\\[\medskipamount]
&=&\int d^NQ\,|M|^{\haf}\tilde\psi^*\left(f^a(Q)\right)
\tilde \chi\left(f^a(Q)\right)\nonumber\\[\medskipamount]
&=&\int d^nq \,d^kq  \left|{\partial Q\over\partial q}\right|
\,|M|^{\haf}\tilde\psi^*\left(q^a\right)
\tilde \chi\left(q^a\right)\nonumber\\[\medskipamount]
&=&\int d^nq \left[\int d^Kq |m|^{\haf}\right]\tilde\psi^*
\left(q^a\right)\tilde
\chi\left(q^a\right)\nonumber\\[\medskipamount]
&=&\int d^nq\,\mu\left(q^a\right)\tilde\psi^*\left(q^a\right)
           \tilde \chi\left(q^a\right). \label{5.16}
\end{eqnarray}
The physical position states $\left|q^a\right>_{\rm Ph}$ can be written
in terms of the position states $\left|Q\right>$ in~$\cal H$.
 Indeed, using (\ref{5.11}) we obtain
\begin{equation}
\left|q^a\right>_{\rm Ph}={1\over\mu\left(q^a\right)}\int d^NQ\,
|M|^{\haf}\delta^n\left(q^a-f^a(Q)\right)\left|Q\right>. \label{5.17}
\end{equation}

In adapted coordinates, using the factorization property, equation (\ref{5.A}),
we obtain
\beq
|q^a\rag_{Ph}=\frac{1}{\cv(q^a)}\; \int\, dq^k\, |m_{\ab}|^{1/2}\,
|q^a,q^\al\rag\,. \label{5.F}
\eeq
This formula (\ref{5.F}) displays the nature of $|q^a\rag_{Ph}$ as a kind of
average over the gauge degrees of freedom of $|q^a,q^\al\rag$, and this
connection will be useful when we compute the propagator in ${\cal H}_p$.
 The projection onto ${\cal H}_p$ can readily be written as
\begin{eqnarray}
{\cal P}&=&\int d^nq\,\mu\left(q^a\right)
\left|q^a\right>_{\!\rm Ph~\,Ph}\!\left<q^a\right|.
\label{5.18}\\[\medskipamount]
\noalign{\noindent\rm This operator can also be expressed in terms of the
states~$\left|Q\right>$:\medskip}
{\cal P} &=&\int d^NQ\,d^NQ'\,|M|^{\haf}|M'|^{\haf}{1\over\mu
\left(f^a(Q)\right)}\delta^n
\left(f^a(Q)-f^a(Q')\right)\left|Q\right>\left<Q'\right|. \label{5.19}
\end{eqnarray}
As expected, $\cal P$ projects states $\left|Q\right>$ into
$\left|q^a\right>_{\rm Ph}$:
\begin{equation}
\cp\left|Q\right>=\left|q^a\right>_{\rm Ph}, \label{5.20}
\end{equation}
where $q^a=f^a(Q)$.
 Finally, we are now in the position of being able to relate the propagator
in $\cal H$ with the propagator in~${\cal H}_p$.
 Since we have a well defined Hamiltonian $\hat H$ in $\cal H$ which, due
to the first class nature---even at the quantum level---of our system,
evolves physical states into physical states, then we can immediately
write down the extension of formula (\ref{5.17}) to the
Heisenberg representation:
\begin{eqnarray}
\left|q^a,t\right>_{\rm Ph}&=&{1\over\cv\left(q^a\right)}
\int dq^k\,|m_{\ab}|^{1/2}\;
|(q^a,q^\al,t\rag\,.  \label{5.C}\\[\medskipamount]
\smash{\vphantom\rangle_{\rm Ph\!}\left<q^a_{2},t_2
\mid q^a_{1},t_1\right>_{\rm Ph}}&=& {1\over\cv\left(q^a_2\right)\cv(q_1^a)}
\;\int\,dq_1^kdq_2^k\;|m_{\ab}(q_2)|^{1/2}|m_{\ab}(q_1)|^{1/2}
\smash{\vphantom\rangle_{}\left<q^a_{2},q_2^\al,t_2
\mid q^a_{1},q_1^\al,t_1\right>}\,. \nonumber \\ %% \label{5.E}
\end{eqnarray}

This equation is one of the main results of this paper.

\section{Quantum First Class Condition and the Gauge Group.}
\setcounter{equation}{0}

In this section we want to explore the possibilities afforded by virtue of
the enormous freedom in the choice of a non-singular $M^{AB}$ satisfying
equation (\ref{3.9}). Its selection should be constrained by at least the
requirement that the first class nature of the system be preserved at the
quantum level. This is essential for the implementation of the Dirac
program. Once this has been achieved, we can then address the issue of
coincidence of Dirac's method with the ``first reduce'' quantization scheme.
It is not entirely clear to us that this is something which should limit
the possible choices of $M^{AB}$, since there doesn't seem to be a physical
principle which would instruct us to favor one method over the other. In
view of this we will keep an open mind on this issue, but always having
present that this is an important aspect of any comparison between
different methods of quantization, since the existence of inequivalent
 quantization
schemes for constraint systems is  something that should not be overlooked
so lightly.

\nd
{\it QUANTUM FIRST CLASS CONDITION}

The expression for $M^{AB}$ in the adapted coordinate system, equation
(\ref{4.12})
$$
M= (m^{A'B'}) = \left(\begin{array}{cc}g^{ab}&m^{a\beta}\\[\medskipamount]
m^{\alpha b}&m^{\alpha\beta}\end{array}\right) %\label{4.12}
$$
allows us to write the Laplace-Beltrami operator(acting on scalars) as
\beq
\Del_M= m^{A'B'} \nabla_{A'}\pa_{B'} = m^{A'B'} (\pa_{A'}\pa_{B'} -
\Gam_{A'B'}^{C'}\pa_{C'})  \label{6.1}
\eeq
or, on physical wave functions:

\beq
\Del_M\mid_{Ph} = g^{ab}\,\pa_a\,\pa_b -m^{A'B'}\Gam_{A'B'}^{a} \pa_a~ .
\label{6.2}
\eeq
The first class condition at the quantum level, equation (\ref{5.5}), is
equivalent to preserving
the physical nature of the states under Hamiltonian evolution, which in
our case, in virtue of the gauge invariance of the potential, equation
(\ref{2.6}), reads
\beq
\hat U_\al\, \Del_M\, |{\rm Phys}\rag =0,~~\al=1,\cdots, k \label{6.3}
\eeq
which leads, considering the form of (\ref{6.2}), to the following sufficient
and necessary condition:
\beq
\hat U_\al\, (m^{A'B'}\Gam_{A'B'}^a) = 0,~\al=1,\cdots, k\,,\, a=1,\ldots,n\,.
\label{6.4} \eeq
In other words, $m^{A'B'}\Gam_{A'B'}^a$ has to be a gauge invariant object.

One criterium that immediately comes to mind for the determination of
$m^{A'B'}$
in order to fulfill equation (\ref{6.4})
is to demand that it be a Killing metric for the generators \break  $\hat
U_{\al,} \;\al=1,\cdots, k$
of the gauge transformations \cite{re:f 9}.
This guarantees that the Hamiltonian $\hat H$ of equation (\ref{5.3})
commutes
with these Killing vectors , which is a strong way to satisfy the
first class condition,equation (\ref{6.4}). For this to  be achieved,
  these
generators have to   form    a Lie algebra, i.e., the structure
functions in (\ref{2.12}), $C_{\ab}^\gam(Q)$,  have to be constant. It
 is easy to
show that this is nothing but the integrability condition for
the existence of a
Killing metric constructed with these generators. Hitherto we
have not had to
mention the freedom to choose a basis of the space of null vectors
of the
Lagrangian
metric $G_{AB}$. This  corresponds to the invariance of the constraint
surface defined by (\ref{3.2}) under rescaling of the primary Hamiltonian
constraints \cite{re:f 10}.  By selecting the generators $\hat
U_{\al,}\,\al=1,\cdots,k$ to   form a Lie algebra (i.e., constant
structure
coefficients), we would be spoiling this rescaling invariance. Moreover,
different ways of selecting the generators may lead to different
Lie algebra
structures, and hence different Killing metrics. This will likely lead to
different quantizations. These reflexions prompt us to discuss the nature of
the
gauge group at more length.

\nd
{\it THE GAUGE GROUP}

At a fixed time, instead of considering the gauge group acting on the
space of trajectories, we can consider its action on $\cq$. Here
the gauge group is defined as the group
of transformations (diffeomorphisms) that leave the orbits invariant.
In adapted local coordinates \cite{re:f 11},
\beqra
&&    \!\!\!\!\!q^a\,,\, q^\al \llra \bar q^{a}\,,\, \bar q^{\al}
\nonumber\\[4pt] \bar q^{{}{a}}\! &=& \!\!\!q^a~,~ \bar q^{\al} = \bar
q^{\al}(q^a\,,\,q^\al) .\label{6.5} \eeqra

\vspace{-36pt}
\nd  with

\vspace{12pt}
\nd
The null vector fields  of the Lagrangian metric, described
by the basis $\hat U_{\al,}\,\al=1,\cdots,k$, are tangent to the
orbits and
generate  uniparametric subgroups of the gauge group. There is a
great deal of
freedom to choose these generators. In particular it may be possible
to choose $k$ independent generators spanning a Lie algebra
structure which will be associated with a $k$-dimensional
Lie subgroup. Locally, one can always obtain these Lie subgroups of the
gauge group, but global obstructions may rule out some of them. One
simple example of such a local construction is furnished by the
holonomic basis  $\pa/\pa q^\al,\al=1,\cdots,k$,
which provides us with an
Abelian  subgroup. These considerations serve to illustrate the following
important point: {\it any of these subgroups can be used to reconstruct
the orbits, and hence, the full gauge group}. So, in this geometrical
picture, there is no loss of information if we ``cut down'' the gauge
group to one of these subgroups \cite{re:f 12}.

The possibility of ``changing'' the structure of
the gauge group (for instance
``Abelianization'' of the gauge group)  has been
dealt with  by the authors of reference  \cite{re:BV 84}.
Nevertheless, in the
light of our previous discussion, we see that this kind of
language can be
somewhat misleading, since one is never changing the structure
of the gauge
group but just selecting different subgroups (or more general
objects as for instance generators that close with non-constant structure
functions) to describe the orbits.

After this discussion on the gauge group, it should be quite clear what
we meant above by the statement that different ways of selecting the
generators  may lead to different Lie subgroups, hence different Killing
metrics, and possibly inequivalent quantizations. It would be interesting  to
find some principle, in the case of an arbitrary  manifold $\cq$,
 that would naturally select one particular Lie subgroup
of the gauge group. At this point we haven't found such a principle, and
it seems to us very unlikely that there is one. Of
course, nothing precludes the existence of such a selection procedure
in some specific cases. Indeed, that will be the case of $\cq=G$
with $G$ a compact group, which will be the subject of the next
section.

If we assume the Killing condition one can try to make contact with the
``reduce first'' method as follows. First of all, as we saw in Section 5,
we have to check for consistency of the dynamical evolution in both
approaches, i.e.,
\beq
\Del_M\mid_{Ph} = \Del_g \label{6.6}
\eeq
where $\Del_g$ is defined by equation (\ref{4.17}). In the adapted coordinate
system, (\ref{6.6}) implies
\beq
m^{A'B'}\Gam_{A'B'}^a = g^{cb}\widetilde\Gam_{cb}^a .\label{6.7}
\eeq
In equation (\ref{6.7}) $\widetilde\Gam_{cb}^a$ is computed with the reduced
metric. This would be readily guaranteed if
the metric $M$ in adapted coordinates reads as follows:
\beq
m^{A'B'}= \left(\begin{array}{cc} g^{ab}(q^a) & 0 \\[\medskipamount]
0 & g^{\ab}(q^\al)\end{array} \right)\,. \label{6.8}
\eeq
Consistency of this form of the metric and the Killing condition demands
that
\beq
U_{\al,a}^\beta = 0 \label{6.9}
\eeq
which, considering the freedom in the choice of the $\hat U_\al$ to form a
Lie algebra does not seem to be, at least locally, too restrictive.

The other issue to study in order to check for matching between the
two approaches is the relation between the measures in both cases.
At this point the factorization property, eq. (\ref{5.A}), is very useful.
Notice that in general the volume of the orbit $\cv(q^a)$ is a function of
$q^a$.  In the case the metric takes the form (6.8) a $q^a$ dependence would
arise if the range of integration in eq. (\ref{5.B}) were to depend on the
orbit which is a global problem that one should not discard a priori.
Equation (\ref{5.B}) shows that in the general case, when $\cv(q^a)$ is not a
constant , there is a good chance
for the two procedures not to agree, since the measures are not equal.
One could attempt to remedy this deficiency by absorbing $\cv(q^a)$
in a redefinition of the wave functions. Nevertheless, this will generally
spoil the equivalence between the dynamics which had been previously
established. In the situation where $\cv$ is a constant, one would be
tempted to conclude that equivalence has been achieved. But there might be
some issues concerning the domain of the operators when the self-adjointness
condition is implemented, which we have taken for granted. This clearly
deserves further study on a case by case basis.

Finally, we would like to make the following remark. The consistency of
the dynamical evolution of the two approaches depend only
on the form of the metric (\ref{6.8}),{\sl
regardless of it being Killing or not}. Moreover, within our
formulation we can always choose $m^{A'B'}$ to be of this form, in virtue of
the
freedom afforded to us and  displayed in equation (\ref{4.12}).  It
now seems more plausible that by playing with coordinate choices
for the orbits
one can obtain volumes which are orbit independent in the case the metric takes
the form (\ref{6.8}). Indeed, if we could
find gauge coordinates with constant range, then all we would do is write
down (\ref{4.12}) in these coordinates in the form (\ref{6.8}).
In this case, again up to the resolution of issues of  domains of the
operators in the theory, we would have achieved equivalence.

At this point the reader may have developed the feeling that, given the
multiplicity of options available in the Dirac quantization approach,
 equivalence with reduced quantization is almost a  matter of chance.
Nevertheless,
there is a wide class of theories where equivalence is rigorously obtained,
with a natural selection of the Lie subgroup of the gauge group and the
metric.
This is the case of quantization on coset spaces of compact Lie groups,
which is the subject of next section.
Before that it is worth to mention that Kucha\v r \cite{re:kk}
as shown in a simple
model the non-equivalence between Dirac and reduced quantization. His
model has a field theoretical application
in scalar electrodynamics but, as it is pointed out by Kucha\v r,
renormalization problems obscure in this case the issue of
the equivalence of both approaches.

\section{Application:  Quantization on Coset Spaces}
\setcounter{equation}{0}

One important example of theories where our previous results can be
applied is the case of quantization on coset spaces of compact Lie
groups.
We will first establish notation and then explain how to set up the
constraint systems of interest for us in such manifolds.

Our configuration manifold will be a compact Lie group $G$ of
dimension~$N$.
Its Lie algebra will be realized by means of left-invariant vector
fields $\hat U_A$, $A=1,\ldots,N$:
\begin{eqnarray}
\left[\hat U_A,\hat U_B\right]&=&C^C_{AB}
      \hat U_C, \label{7.1}\\[\medskipamount]
\noalign{\noindent\rm where the structure constants satisfy, besides the
obvious ones, the following antisymmetry property:\medskip}
C^C_{AB}&=&-C^B_{AC} \label{7.2}
\end{eqnarray}
which can always be achieved for any compact Lie algebra.

On coordinates $Q^A$, $A=1,\ldots,N$ our left-invariant vector fields
can be written as $\hat U_A=U^B_A(Q)\partial/\pa \,Q^B$, and its dual
left-invariant forms as $\Om^A=\Om^A_B(Q)\,dQ^B$, $\Om_A^B = (U^B_A)^{-1}$.
 $G$~is endowed with a  non-singular left-invariant metric
$M_{AB}$ which is a Killing metric for the vector fields
$\hat U_A$, $A=1,\ldots,N$.
It has the form
\begin{equation}
M_{AB}=\Om^C_A\Om^D_B\delta_{CD}. \label{7.3}
\end{equation}
In order to introduce a constraint system in a natural fashion, we will
extract from (\ref{7.3}) a singular metric which will define the kinetic
term in the Lagrangian for such a system.
One way to achieve this is simply to restrict the range of the summation
index in (\ref{7.3}) to $a=1,\ldots,n<N$.
Indeed, $\hat U_\alpha$, where $\alpha$ runs over the complementary set
of indices, form a basis of null vectors for this new metric
$G_{AB}=\Om^a_A\Om^a_B$ \cite{re:f 13}.
 At this point, we start making contact with our
Lagrangian setting of Section~2.
In this case, condition (\ref{2.8}) implies that the structure constants
have to satisfy
\begin{eqnarray}
C^a_{\alpha\beta}&=&0 \,,\label{7.4}\\[\medskipamount]
\noalign{\noindent\rm and\medskip}
C^a_{b\alpha}&=&-C^b_{a\alpha}\,. \label{7.5}
\end{eqnarray}
Equation~(\ref{7.4}) is the statement that the $\hat U_\alpha$,
$\alpha=1,\ldots,k=N-n$ form a subalgebra of the original Lie algebra:
\begin{equation}
\left[\hat U_\alpha,\hat U_\beta\right]=
C^\gamma_{\alpha\beta}\hat U_\gamma. \label{7.6}
\end{equation}
Equation~(\ref{7.5}) does not give new information, since this already
guaranteed by (\ref{7.2}).

This subalgebra generates a Lie subgroup $K$ of~$G$.
The quotient manifold $\cal M$ of Section~4 is then the coset space
$G/K$.
 To fix ideas, it is convenient to write our singular Lagrangian:
\begin{equation}
L={1\over 2}G_{AB}\dot Q^A\dot Q^B-V, \label{7.7}
\end{equation}
where $G_{AB}$ was just defined above, and $V$ is a potential function
satisfying, according to (\ref{2.6}), $U^A_\alpha V_{,A}=0$,
$\alpha=1,\ldots ,k$.

Now we present the Hamiltonian formalism.
According to (\ref{3.2}) the constraints are
\begin{equation}
\varphi_\alpha\equiv U^A_\alpha(Q)P_A\approx 0,
\qquad\alpha=1,\ldots,k\,. \label{7.8}
\end{equation}
Due to the particular structure of our metric $G_{AB}$ it is
straightforward to write down a solution to equation~(\ref{3.9}),
namely, $U^A_aU^B_a$, which is obviously singular.
The freedom of choice of $M$ as shown in equation~(\ref{3.10}) allows us
to work with a non-singular contravariant metric
$M^{AB}= U^A_CU^B_C$.
This is nothing but the inverse of the covariant non-singular metric
$M_{AB}$ that we started with.
This metric structure provides us with the invariant measure on $G$, and
it is such a natural structure on it, that we will henceforth work
always with it.  This illustrates our point that the freedom in the choice of
$M$ can be of great importance.
Therefore, (\ref{3.7}) becomes
\begin{equation}
H={1\over 2}~M^{AB}P_AP_B+V\,, \label{7.9}
\end{equation}
with $M^{AB} =U^A_C \,U^B_C$.
Now we have all the ingredients to quantize according to the two schemes
presented in Sections~4 and~5.
The ``first reduce and then quantize'' procedure is readily implemented
since all we have to do is use the metric tensor $M$ to obtain
$\tilde g$ for $G/K$ (equation~(\ref{4.6})).
This assignment defines the Hilbert space structure and the quantum
dynamics according to \mbox{(\ref{4.15})--(\ref{4.18})}.
This completes the ``first reduce~\ldots\,'' program.

In order to implement the ``first quantize and then reduce'' scheme, we
first have to check the first class character of the quantum
Hamiltonian, i.e., equation~(\ref{5.5}).
The first class nature of the constraints is already assured by
(\ref{7.6}).
In our case the Laplacian operator is
\begin{equation}
\Delta_M=|M|^{-\haf}{\partial\over\partial Q^A}|M|^{\haf}
             M^{AB}{\partial\over\partial Q^B}. \label{7.10}
\end{equation}
Now, since the $\hat U_A$, $A=1,\ldots,N$ are Killing vectors for
$M$, they are divergenceless:
\begin{equation}
|M|^{-\haf}{\partial\over\partial Q^A}
|M|U^A_C=0,\qquad C=1,\ldots,N \label{7.11}
\end{equation}
which allows us to write $\Delta_M$ as
\begin{equation}
\Delta_M=\hat U_C\hat U_C. \label{7.12}
\end{equation}
Using equation (\ref{7.12}) and the antisymmetry property
$C^C_{AB}=-C^B_{AC}$ is a matter of straightforward algebra to verify
that \cite{re:f 14}
\begin{equation}
\left[\Delta_p,\hat U_A\right]=0,\qquad
A=1, \ldots ,N\,. \label{7.13}
\end{equation}
Hence, $\hat H$ passes the test.
This guarantees a consistent quantization in ${\cal H}_p$, in the sense
defined in Section~5.
Furthermore, since the gauge orbits (generated by~$K$) are compact, all
the formulae derived in that section apply in a rigorous sense.
We now proceed to show equivalence between the two approaches.
This will be done in two steps: first, we will show that the measure on the
physical space $\cah_p$ coincides, up to an  irrelevant constant factor, with
the measure obtained from the ``reduced first''   procedure - this shows that,
in fact, the two Hilbert spaces are the same . Then, we will see that the
dynamics in both cases are identical.

\vspace{12pt}
\nd
{\it THE MEASURES.}

In the ``first reduce'' procedure, the metric in $G$ is defined as
\beqra
g^{ab} & = & M^{AB}\,\frac{\pa q^a}{\pa Q^A}~\frac{\pa q^b}{\pa Q^B}
= U_D^A\;U^B_D \; \frac{\pa q^a}{\pa Q^A}~\frac{\pa q^b}{\pa Q^B}=
\nonumber\\[4pt]
&=& \hat U_D(q^a)\hat U_D(q^b) =\hat U_d(q^a)\hat U_d(q^b) =
U^a_d\,U^b_d \label{7.14}
\eeqra
 where we used $\hat U_\al(q^a)=0$.  The measure is then
\beq
\mu_R = |g|^{\haf}= |g^{ab}|^{-\haf} = |U_a^b| ^{-1} \label{7.15}
\eeq
Now consider the ``first quantize'' approach. In the adapted coordinate
 system, $ Q^{A'}= (q^a,q^\al)$,
we showed factorization in general, eq. (\ref{5.A}).  The only possible
obstruction to equivalence is the dependence of $\cv(q^a)$ on $q^a$.  Let us
now compute $\cv(q^a)$ in this case.  From (\ref{7.3}) we get
\begin{eqnarray}
|m|^{1/2} &=& | \hat U_D(Q^{A'})|^{-1} =  |U_a^b|^{-1}\, |U_\al^\beta|^{-1}
\nonumber \\[4pt]
&=& \mu_R(q^a)|U_\al^\beta|^{-1} = \mu_R(q^a)|\Om_\al^\beta|\eqv
\mu_R(q^a)|\Om| \label{7.A}
\end{eqnarray}
Therefore, the volume of the orbit is:
\beq
\cv(q^a) = \int\, d^kq\,|\Om|\,.  \label{7.B}
\eeq
We now prove that (\ref{7.B}) is indeed a constant.
To see this, it is convenient to rewrite $\cv$ in intrinsic notation.
Consider the injection
\def\noteps{\hbox{{$\eps$}\kern-.45em\hbox{/}}}
\beq
i_{q^a}~:~\cO(q^a) \llra G\,. \label{7.22}
\eeq
Then we have
\beq
\cv(q^a) = \int_{\cO(q^a)} i^*_{qa}{\textstyle\sum}\,, \label{7.23}
\eeq
where $\sum=\Om^1\wedge\Om^2\wedge\cdots\wedge\Om^k$  and the indices
$1,2,\cdots k$ are the $\al$-type
indices. $i^*_{q^a}\sum$ is the pullback of $\sum$ under
 (\ref{7.22}). We should remind ourselves that $\sum$ is by construction
left-invariant. Now consider an element $g\,\eps\,G$, but $g\,\noteps\,K$. It
maps orbits into other orbits
\beq
g~:~\cO(q^a) \llra \cO(\til q^{a}) \label{7.24}
\eeq
Then the following chain of equalities holds:

\def\raiseo{\raise .2mm\hbox{\small$\circ$}\normalsize}
\beq
\cv (\til q^{a}) \eqv \int_{\cO(\til q^{ a})} i^*_{q^a}{\textstyle\sum} =
\int_{\cO(q^{ a})} g^*( i^*_{\til q^a}{\textstyle\sum}) = \int_{\cO(q^a)}
i_{q^a}^*(g^*{\textstyle\sum})=\int_{\cO(q^a)} i^*_{q^a} {\textstyle\sum}
\eqv\cv (q^a) .\label{7.25}
\eeq
The first equality comes from a passive interpretation of the action
of $g$ as a change of variables in the same orbit. The second one is
just a consequence of the fact that \break
$i_{\til qa~\raiseo}g=g_{\raiseo}\,i_{qa}$. To obtain the last one we used the
left-invariance of $\sum$. This proves that all orbits have the same volume.
Therefore  the measures are equal up to a normalization constant.

\vspace{6pt}
\nd
{\it THE DYNAMICS}

Now we proceed to show that the Laplacian operator (\ref{7.10}), $\Del_M$,
 when restricted to $\cah_p$  coincides with the Laplacian operator
(\ref{4.17})  $\Del_g$ of the ``reduce first'' approach. First we observe that
(\ref{7.12}) allows us to write the restriction of $\Del_M$ on $\cah_p$ as
\beq
\Del_M\mid_{\cah_p} = \hat U_a\, \hat U_a, \label{7.26}
\eeq
where we have used $\hat U_\al q^a=0$. Explicitly, in the adapted coordinate
system,
\beqra
\Del_M\mid_{\cah_p} &=& U_a^b\frac{\pa}{\pa q^b}
U_a^c\;\frac{\pa}{\pa q^c} + U_a^\al\,
\frac{\pa}{\pa q^\al}\;U_a^c\,\frac{\pa}{\pa q^c}\,\mid _{\cah_p}
\nonumber\\[4pt] &=& U_a^b\,\frac{\pa}{\pa q^b}\;U_a^c\;\frac{\pa}{\pa q^c} +
U_a^\al\,U_{a,\al}^c \;\frac{\pa}{\pa q^c}. \label{7.27}
\eeqra
On the other hand,
\beqra
\Del_g &=& |g|^{-\haf}\,\pa_b\,|g|^{\haf}\,\,g^{bc} \,\frac{\pa}{\pa q^c} =
|g|^{-\haf}\,\pa_b |g|^{\haf}\,U_a^b\,U_a^c\;\frac{\pa}{\pa q^c}
=\nonumber\\[4pt]
 &=& U_a^b\,\frac{\pa}{\pa q^b}\,U_a^c\;\frac{\pa}{\pa q^c} +
|g|^{-\haf}\, \Big(|g|^{\haf}\,U_a^b\Big)_{,b}\,U_a^c\;\frac{\pa}{\pa
q^c}\,.\label{7.28} \eeqra
Hence
\beqra
\Del_M\mid_{\cah_p}-\Del_g &=& \Big( U_a^\al\,U_{a,\al}^c -
|g|^{-\haf}\,\big(|g|^{\haf}\,U_a^b\big)_{,b}\,U_a^c\Big) \,   \frac{\pa}{\pa
q^c} \nonumber \\[4pt]
& \eqv & A^c\;\pa_c \,.\label{7.29}
\eeqra

We are going to show that $A^c=0$. Using the fact that $|g|_{,\al}=0$, we can
write
\beq
|g|^{-\haf}\,\Big(|g|^{\haf}\,U_a^b \Big)_{,b} =|g|^{-\haf}\,
\Big(|g|^{\haf}\,U_a^{A'} \Big)_{,A'} - U_{a,\al}^\al \label{7.31}
\eeq
where $A'$ runs over the entire set of indices $a$ and $\al$.
Then
\beqra
A^c &=& U_a^\al\; U_{a,\al}^c - \left(|g|^{-\haf}\Big(|g|^{\haf}\,
U_a^{A'}\Big)_{,A'} - U_{a,\al}^\al\right) U_a^c \nonumber \\[4pt]
&=& (U_a^\al\, U_a^c)_{,\al} - |g|^{-\haf}\Big(|g|^{\haf}\,
U_a^{A'}\Big)_{,A'}\, U_a^c\, .\label{7.32}
\eeqra
The divergenceless of the vector field $\hat U_a$ can be reexpressed as
\beqra
0 &=& |m|^{-\haf}\, \Big( |m|^{\haf}\,U_a^{A'}\Big)_{,A'} =
|g|^{-\haf}\,|\Om|^{-1}\Big(|g|^{\haf}|\Om|\, U_a^{A'}\Big)_{,A'} \nonumber
\\[4pt] &=& |\Om|^{-1}\, |\Om|_{,A'}\, U_a^{A'} + |g|^{-\haf}
\Big( |g|^{\haf}\,U_a^{A'}\Big)_{,A'}\,, \label{7.33}
\eeqra
which allows us to rewrite (\ref{7.32}) as
\beq
A^c = (U_a^\al\, U_a^c)_{\,,\al} + |\Om|^{-1}\, \hat U_a\Big(|\Om|\Big)\,
U_a^c.\label{7.34}
\eeq
Now, recall that $\hat U_\al = U_\al^\beta\, \pa_\beta~,~\pa_\al =
\Om_\al^\beta\, \hat U_\beta$. Also,
\beqra
|\Om|^{-1}\, \hat U_a\Big(|\Om|\Big) &=& |\Om|^{-1}\, |\Om|\, \Big(
-\Om_\al^\beta\, \hat U_a(U_\beta^\al)\Big) \nonumber \\[4pt]
&=&- \Om_\al^\beta\, \hat U_a(U_\beta^\al)\,.  \label{7.35}
\eeqra
Using these results (\ref{7.34}) can be rewritten as
\beqra
A^c &=& \Om_\al^\beta \Big( \hat U_\beta(U_a^\al\, U_a^c) -\hat
U_a(U_\beta^\al)
U_a^c\Big) \nonumber \\[4pt]
&=&  \Om_\al^\beta \left( \Big( (\hat U_\beta(U_a^\al) - \hat
U_a(U_\beta^\al)\Big) U_a^c + U_a^\al\, \hat U_\beta(U_a^c)\right).\label{7.36}
\eeqra
 Recalling that $C_{\beta a}^\gam =-C _{\beta\gam}^a=0$ (compactness and
subgroup conditions) we have
\beq
[\hat U_\beta\,, \hat U_a] = C_{\beta a}^d\, \hat U_d\,, \label{7.37}
\eeq
which implies (using $U_\beta^b = 0$) that
\beq
\hat U_\beta(U_a^\al) - \hat U_a(U_\beta^\al) = C_{\beta a}^d\, U_d^\al\,,
\label{7.38}
\eeq
and
\beq
\hat U_\beta (U_a^c)  =   C_{\beta a}^d\, U_d^c\,.
\label{7.39}
\eeq
Hence
\beq
A^c = \Om_\al^\beta\, C_{\beta a}^d (U_d^\al\, U_a^c + U_a^\al \, U_d^c),
\label{7.40}
\eeq
which is zero in virtue of the antisymmetry of the structure constants.
Therefore,
\beq
\Del_M\mid_{\cah_p} = \Del_g . \label{7.41}
\eeq
We would like to emphasize the importance of the compactness of $G$
in the derivation of (\ref{7.41}).

\newpage

\section{Example: The Free Propagator on $S^2$ from The Free Propagator on
$SU(2)$}
\setcounter{equation}{0}

As a concrete application
of the formalism developed in Sections 5 and 7 we will now consider the motion
of a free particle  on a sphere $S^2$, viewed
as the coset space  $SU(2)/U(1)$ \cite{re:f 15}.

Making use of the isomorphism $SU(2)\sim S^3$
we may parametrize the group manifold $SU(2)$
 using polar coordinates on $S^3$:
\beq
    Q^A = (\tha,\phi,\psi),~~ 0\leq\tha\leq\pi,~~0\leq\phi<2\pi,~~0\leq\psi
<4\pi\;.  \label{8.1}
\eeq
The left-invariant vector fields are:
\beqra
\hat{U}_1 &=& \left(\sin\psi\pa_\tha - \frac{\cos\psi}{\sin \tha} \pa_\phi +
\frac{\cos\psi}{\tan\tha} \,\pa_\psi\right) \eqv U_1 \,{}^B\pa_B,
\nonumber\\[6pt]
\hat U_2 &=& \left(\cos\psi\pa_\tha
+\frac{\sin\psi}{\sin\tha}\,\pa_\phi -  \frac{\sin\psi}{\tan\tha}\pa_\psi
\right) \eqv U_2{}^B\pa_B,\\[6pt]
\hat U_3 &=& \pa_\psi\eqv U_3{}^B\pa_B\,,\nonumber \label{8.2}
\eeqra
where $\pa_B\eqv\frac{\pa}{\pa Q^B}$. They satisfy the $SU(2)$ algebra
\beq
[\hat U_A, \;\hat U_B]  =  \eps_{ABC} \hat U_C \;.\label{8.3}
\eeq

The $\hat U_A,\; A =1,2,3$,   are Killing vectors for the metric
\beq
G^{AB}=U_C^A U_D^B\,\del^{CD} \,,   \label{8.4}
\eeq
which in this coordinate system takes the form
\beq
G^{AB} =\left( \begin{array}{ccc} 1 & 0 & 0\\[6pt]   0 & \frac{1}{\sin^2\tha} &
\frac{-\cos\tha}{\sin^2\tha} \\[6pt] 0 & \frac{-\cos\tha}{\sin^2\tha} &
 \frac{1}{\sin^2\tha}\end{array} \right)\;. \label{8.5}
\eeq
We will choose  the uniparametric  subgroup $U(1)$ generated by $\hat U_3$
 as the gauge group $K$.
 The simplest choice of the adapted system is then
\beqra
q^a &=& (\tha,\phi) \nonumber\\
q^\al& = &(\psi)   \;.\label{8.6}
\eeqra

In this system, the measure, equation (\ref{5.14}),  becomes
\beq
 \mu(\tha,\phi) = 4\pi\sin\tha\;,
\label{8.7}
\eeq
and  the propagator, equation (5.25),  now reads
\beq
\phantom{\rag}_{\rm Ph} \lag\tha',\phi',t'|\tha,\phi,t\rag_{\rm Ph}
=\frac{1}{16\pi^2}\,\int_0^{4\pi}d\psi_2 \int_0^{4\pi}d\psi_1
\lag\tha',\phi',\psi_2,t'| \,\tha,\phi, \psi_1, t\rag~. \label{8.8}
\eeq
All we have to do now is
 to substitute for $\lag\tha',\phi',\psi_2,t'|\tha,\phi,\psi_1,t\rag$
its explicit form, which can be readily found
 in the literature. Following Schulman \cite{re:f 16},~ \cite{re:LS 81},
\beq
\lag\tha',\phi',\psi_2,t'|\tha,\phi,\psi_1,t\rag = \frac{1}{16\pi^2}\;
\frac{1}{\sin\,\Gam}\;\sum_{j=0,\haf,1,\cdots} (2j+1)(\sin(2j+1)\Gam)
e^{-\frac{i}{2I}\,j(j+1)(t'-t)}\label{8.9}
\eeq
where $\Gam$ is the geodesic distance between two points on $S^3$:
\beq
\cos\Gam = \cos\left(\frac{\tha}{2}\right)\cos
\left(\frac{\tha'}{2}\right)\cos(\psi_+ -\psi'_+ ) + \sin\left(
\frac{\tha}{2}\right) \sin \left( \frac{\tha'}{2}\right)
\,\cos\,\left( \psi_--\psi'_-\right),  \label{8.10}
\eeq
with $\psi_{\pm}=(\psi{\pm}\phi)/2$, and $\psi'=\psi_2\;,\;\psi=\psi_1$. We use
here notation of reference [37].
Notice we are taking advantage of the isomorphism
 between $SU(2)$ and $S^3$. Performing the integrations in (\ref{8.8}) we then
 obtain
\beq
{}~_{\rm Ph}\lag\tha',\phi',t'| \,\tha,\phi,t\rag_{\rm Ph} =
\frac{1}{16\pi^2}\,
\sum_{j=0}^\infty\; (2j+1) P_j(\cos\,\gam)\, e^{-\frac{i}{2I}\cdot
j(j+1)(t'-t)}\label{8.11}
\eeq
where $\gam$ is the geodesic distance between $q'=(\tha',\phi')$ and
$q=(\tha,\phi)$ on $S^2$. This is in agreement with \cite{re:BJ 87},
after taking
into account the normalizations  for our states, given by equations
(\ref{5.15})
and  (\ref{8.7}).

\newpage

\section{ Conclusions and Outlook}

   There are several lessons that can be drawn from this work, both at a
conceptual as well as at a more technical level, keeping in mind that
we are considering only systems with Hamiltonian and constraints quadratic
and linear in the momenta respectively. On the one hand, the
novelty of the Lagrangian setting for first class systems reveals quite
clearly a source of ambiguities present in the framework of Dirac's
quantization which, to our knowledge, had not been pointed out before.
Within our scheme the one-to-one correspondence between the original
Lagrangian and the reduced Hamiltonian quantization is neatly seen.
Hence the reduced quantization possesses a certain uniqueness that is lacking
in the Dirac approach. This point of view is reinforced  when one considers
our work with path integrals \cite{re:OP} where again
reduced quantization plays a central role. The discussion in section 6 shows
that the
ambiguities mentioned above make plausible the selection of a specific
Dirac quantization which coincides with the reduced first method. Several
features of the gauge group relevant to this problem are also mentioned there.

   On a more practical level we show rigorously that in the case of coset
spaces of compact Lie groups there is a natural selection of Dirac quantization
that fully coincides with the reduced quantization. This fact allows us
to use results of the quantum theory on the group $G$ to obtain the quantum
theory on the coset space $G/K$, with $K$ a subgroup. To make contact with
work of other people we compute the propagator on a two dimensional sphere
$S^2 = SU(2)/U(1)$ by this means.

It would be interesting to extend our work to investigate similar problems in
more sophisticated cases like Chern-Simons which we mentioned in the
introduction. The case of non-compact Lie groups clearly demands an extension
of
our methods in this direction also. We feel   that a close comparison with our
path-integral work
 would be illuminating. Finally, the treatment of
constraints quadratic in the momenta -relevant in the  study of
reparameterization invariant systems like gravity- within our  framework,
deserves our full consideration. Work on these topics is currently in progress.

\vspace{24pt}
\nd
{\Large{\bf Acknowledgments     }}

We are grateful to L.C. Shepley for a useful conversation regarding Killing
vectors.  This work was partly inspired by questions posed to us by J.
Polchinski.  C.R.O. would like to express his gratitude to the World Laboratory
for its generous support, and to the Guggenheim Foundation for support during
the initial stages of this work.  J.M.P. would like to thank the Center for
Relativity of The University of Texas for its warm hospitality.  He also
acknowledges the Ministerio de Educaci\'on y Ciencia from Spain for a grant.
This research was supported in part by the Robert A. Welch Foundation and NSF
Grant PHY 9009850 and NSF Grant PHY 8806567.

\pagebreak

\renewcommand\theequation{A.\arabic{equation}}
\setcounter{equation}{0}

\section*{ Appendix}

\vspace{12pt}
We want to show that from the Hamiltonian (\ref{3.5}),
\beq
H=\haf \;M^{AB}(Q)\, P_A\,P_B ~+~V \label{A.1}
\eeq
and the primary constraints (\ref{3.2}),
\beq
U_\al^A(Q)\,P_A \approx 0, \label{A.2}
\eeq
we can recover the Lagrangian (\ref{2.1}) which
we started with. For this purpose we will perform the inverse Legendre
transformation \cite{re:f 17} on the Dirac Hamiltonian:
\beq
 H_D = \haf\; M^{AB}   P_A\,P_B  ~+~\la^\al\,U_\al^AP_A+V\,.
\label{A.3}
\eeq
This means we want to find $(P_A,\la^\al)$ in terms of
$(Q^A,\dot Q^A)$, as the solution of the following algebraic system of
equations
\beqra
& & \dot Q^A = \{Q^A,\;H_D\} = M^{AB} \,P_B~+~\la^\al U_\al^A
\nonumber \\[4pt]
&  &U_\al^A P_A =0 \,. \label{A.4}
\eeqra
Once this is achieved we will just substitute the momenta as functions
of velocity space variables in the expression for the Lagrangian
\beq
L= P_A\dot Q^A - H\,.\label{A.5}
\eeq
Notice that equation (\ref{A.4}) involves only half of Hamilton equations. This
is so, because this is the one that contains the information about the
Legendre map. In virtue of equation (\ref{A.4}),
\beq
\dot Q^A\, U_{\beta A} = \la^\al \Tha_{\ab} \label{ A.6}
\eeq
where
\beq
\Tha_{\ab}=U_\al^A
U_{\beta A} \eqv
U_\al^A M_{AB} U_\beta^B     \label{ A.7}
\eeq
is just the scalar product of the vector fields $\hat U_\al,\;\hat U_\beta$
with
respect to the metric  $M$. Since the vector fields $\hat U_\al,\;\al=1,\ldots,
k$ are independent,
 and the metric $M$ is non-singular, the matrix $\Tha_{\ab}$ is invertible.
Hence,
\beq
\la^\al = \Tha^{\ab}\; U_{\beta A} \dot Q^A\,, \label{A.8}
\eeq
and using again equation (\ref{A.4}) we obtain
\beq
\dot Q^A = M^{AB} P_B ~+~ \Tha^{\ab}U_{\beta C}\dot Q^C\, U_\al^A \label{A.9}
\eeq
from which we get for $P_B$
\beq
P_B=M_{BA} \cp_C^A\, \dot Q^C , \label{ A.10}
\eeq
where
\beq
\cp_C^A = \del_C^A - \Tha^{\al \beta} U_{\beta C} U_\al^A \label{A.11}
\eeq
is the projector in the direction transverse to the orbits. Finally,
the Lagrangian becomes
\beq
L = \haf\;M_{CD}^\bot \dot Q^C\dot Q^D - V, \label{ A.12}
\eeq
where
\beq
M_{CD}^\bot = \cp_C^A\, M_{AB}\, \cp_D^B\,. \label{ A.13}
\eeq
It only remains to show that $M_{CD}^\bot =G_{CD}$, where $G_{CD}$ is
the metric in the original Lagrangian (\ref{2.1}). To show this it is
convenient
to work in the adapted coordinate system, in which the obvious  fact that the
vectors $\hat U_{\al,}\al=1,\ldots, k$ are null vectors of $M_{CD}^\bot$
reveals  that it has the  form
\beq
M^\bot=\left( \begin{array}{cc} M^\bot_{ab} & 0\\[3pt]
0&0 \end{array}\right)\,. \label{A.14}
\eeq
In view of equation (\ref{4.12}), in order to check that $M_{ab}^\bot = g_{ab}$
we only need to verify  $M^{aA'}M_{A'b}^\bot = \del_b^a$. From the definition
of
$M^\bot$ we obtain
\beq
M_{A'B'}^\bot = M_{A'B'} - M_{A'\rho}U_\al^\rho\Tha^{\ab} U_\beta^\si M_{\si
B'}\,, \label{A.15}
\eeq
where  $U_\al^a = 0, $ equation (\ref{4.9}), was used. Finally, recalling that
$\del_\rho^a=0$ we have
\beq
M^{aA'} M_{A'b}^\bot = \del_b^a -\del_\rho^a U_\al^\rho\Tha^{\ab} U_\beta^\si
M_{\si b} = \del_b^a\,, \label{A.16}
\eeq
which proves our assertion.


\begin{thebibliography}{99}
\bibitem{re:f 1} This has been known [4]
for some time, but still remains an open problem.
\bibitem{re:f 2}  Throughout this paper we will consider explicitly the case of
a quantum mechanical system with a finite number of degrees of freedom.
Nevertheless, our results will readily generalize to the case of a field
theory, once DeWitt's condensed notation is adopted.
The important application to Yang-Mills fields requires a delicate
analysis of the role played by the time component $A_0$ as a Lagrange
multiplier.  In this way we can get rid of the canonical
momenta conjugate to $A_0$  and obtain
a non-covariant formulation of
Yang-Mills theories with only primary constraints, the Gauss law, which
were secondary constraints in the covariant formulation.
Once this is done, our results can be applied.
\bibitem{re:f 3} We use the convention that a comma denotes partial derivative.
\bibitem{re:f 5}  This result is a direct consequence of the analysis of the
relationship between Hamiltonian and Lagrangian constraints, which was
performed
by the authors of reference [22].  There it was shown that at every
stage of the stabilization algorithm (i.e., the iterative procedure of
generating the constraints of the theory, which appear by consistency
requirements) the Lagrangian constraints can be obtained by means of either of
the following two constructions:~~a)~~ Some will come as the pullback of the
secondary, terciary,\ldots, constraints which arise from the Hamiltonian
version
of the algorithm mentioned above;~~b)~~ The others will come  as the Lagrange
multipliers in the Dirac Hamiltonian are determined as specific functions in
phase space.  Having a first class Hamiltonian system without secondary
constraints (and hence without further constraints) renders these two
constructions meaningless. Hence there will be no room for Lagrangian
constraints.  For more details we advise the reader to consult the reference
given above.
\bibitem{re:f 6}  We are using here free-index notation.  Let $G$ be
the convariant metric tensor $G_{AB}\,dQ^A\otimes dQ^B$.  Then
$i_{\hat U}G=G_{AB}U^AdQ^B$.  Hence $i_{\hat U}G=0\Leftrightarrow G_{AB}U^A=0$.
\bibitem{re:f 7}  Notice that $\al$ runs from $n+1$ to $N$.
Nevertheless, to simplify notation, we will have it run from $g$ to $k=N-n$.
Consider the following identity
\bibitem{re:f 8}  From here on, we will use the
convention that $\int d^nq$ and $\int d^kq$ denote integration over variables
$q^a$ and $q^\al$ respectively.
\bibitem{re:f 9} In usual treatments which ignore
the  ambiguities present in the Hamiltonian kinetic term  of a constraint
system
,
this condition on the metric is taken as the starting point.
\bibitem{re:f 11}  Since we are not specifying the
manifold $\cq$ at this point, we will ignore global considerations
which may arise. Everything we do here is valid locally.
\bibitem{re:f 12}  Very interesting comments along these lines were expressed
by
B. DeWitt in [33].
\bibitem{re:f 13}   We are here following the
convention adopted in [25].  Repeated indices run over their
corresponding range.
\bibitem{re:f 14}  Equation (\ref{7.13}) is already guaranteed by the Killing
vector condition.
\bibitem{re:f 15} This problem has recently been
studied by B.P.Dolan [37] within the BRST approach.Here we show that
full equivalence between the ``reduce  first'' and the Dirac approaches can be
achieved in this case(coset space) without having to introduce ghosts.
\bibitem{re:f 16} Notice that we are introducing
 the ``moment of inertia" $ I$ as used in equation (\ref{4.17}) of reference
[39]. This makes comparisons more tractable.
\end{thebibliography}
\end{document}